\documentstyle [psfig,11pt]{article}
\def\l {\lambda } 
\def \t {\theta }

\def\a {\alpha }

\def \d {\delta }
\def \D {\Delta }

\def \g {\gamma }
\def \G {\Gamma }

\def \b {\beta }

\def \s {\sigma }
\def \e {\epsilon }
\def \ud { {1 \over 2} }

\def \qslash {Q \kern -.5em\slash }
\def \pslash {p \kern -.5em\slash }
\def \ppslash {p' \kern -.5em\slash }
\def \Pslash {P \kern -.5em\slash }
\def \kslash {k \kern -.5em\slash }
\def \bea {\begin{equation}}
\def \eea {\end{equation}}

\def \pr  { Phys. Rev. }
\def \np { Nucl. Phys. }

\def \prl { Phys. Rev. Lett. } 
\def \pl { Phys. Lett. }

\input psfig
\if@twoside\oddsidemargin25mm
\else\oddsidemargin25mm\evensidemargin25mm\marginparwidth25mm\fi
\begin{document}
\begin{titlepage}
\title{CP violation  flavor  asymmetries  in slepton pair
production at leptonic  colliders from broken R parity \thanks 
{\it Supported by the Laboratoire de la Direction des Sciences
de la Mati\`ere du Commissariat \`a l'Energie Atomique}}
\author{{M. Chemtob ,  G. Moreau} \\ \\
{\em  Service de Physique Th\'eorique}\\
{\em CE-Saclay F-91191 Gif-sur-Yvette, Cedex FRANCE}}
\date{\today }
\maketitle

\begin{abstract}
We examine the effect of the R parity odd, lepton number violating,
renormalizable interactions on flavor non-diagonal rates  and CP asymmetries  
in the production of slepton pairs,  
$e^-+e^+\to \tilde e_{HJ} +\tilde  e^\star _{H'J'}, \ [J\ne J'] ,\ [H,H'=(L,R)] $
at leptonic colliders. The R parity odd coupling constants are
assumed to incorporate CP odd complex phases. 
The flavor changing rates are controlled  by tree level amplitudes 
and quadratic products of different R parity violating coupling constants 
and  the  CP violating asymmetries by interference terms between tree and loop level  amplitudes and quartic products. The consideration  
of  loop amplitudes  is restricted  to the 
photon and  Z-boson vertex corrections.
We present numerical  results  using a
family and (quarks and leptons)  species
independent mass parameter, $\tilde m$, 
for all the scalar superpartners and making simple assumptions for
the family dependence of the R parity odd coupling constants.
The flavor non-diagonal rates, $\s_{JJ'}$, vary in the range, $({\l \over 0.1})^4 \  2\ - \ 20 $ fbarns, for sleptons masses $\tilde m < 400 $ GeV, 
as one spans the interval of center of mass  energies from 
the Z-boson  pole up to $ 1000$ GeV. For sleptons masses,  
$ \tilde m> 150 $ GeV, these observables  could be of use at NLC energies 
to set useful bounds on the R parity odd coupling constants. 
The  predicted asymmetries are in order of magnitude,
${\cal A}_{JJ'}= { \s_{JJ'} -\s_{J'J}
\over \s_{JJ'} +\s_{J'J}  } \simeq 10 ^{-2}  \ - 10^{-3}$. 

\end{abstract}


Saclay {\bf T98/062}

{\it hep-ph/9808428}

For correspondence: {\it  M. Chemtob:  Address:
Service de Physique Th\'eorique, 
CE-Saclay F-91191 Gif-sur-Yvette, Cedex FRANCE; 
Telephone: 01 69 08 72 65; 
Fax: 01 69 08 81 20; Email: chemtob@spht.saclay.cea.fr}
\end{titlepage}

\section{Introduction}
\label{secs1}
On side of the familiar low  energy tests of CP symmetry non-conservation,
a large number of tests have been developed  over the years 
for high energy colliders \cite{cpcoll1,cpcoll2,cpcoll3}. 
The existing proposals have  dealt with  different types of CP odd
observables (quark and leptons flavor aymmetries \cite{cpobs0},
spin polarization asymmetries \cite{cpobs1,cpobs2,cpobs3}, 
heavy quarks or leptons electric dipole moments \cite{wermes}, ...)
and covered a wide variety of physical processes, ranging from  decay reactions 
($Z, \  W^\pm $ gauge 
bosons \cite{cpobs0,bern1}, Higgs bosons  \cite{chang1,grzad1} 
or  top-quarks \cite{grzad2})
to production reactions (leptons-antileptons  and light 
quarks-antiquarks  pairs \cite{cpobs0}, single top-quarks \cite{atwood1}, top-antitop-quark pairs \cite{peskin1,bartl,baek}, or superpartners pairs, 
$\tilde \chi^+ \tilde \chi^-, $ \cite{kizu} $ \tilde q \tilde {\bar q} ,$
\cite{pilaf}  and  
$\tilde l^+ \tilde  l^ -$ \cite{arkani2,bowser}).
For lack of space, we have  referred to those works 
from which one could hopefully trace the extensive  published literature.

One of the primary motivations for these high energy tests 
 is the search for physics beyond the standard model. The supersymmetry 
option is especially attractive in this respect since any slight generalization 
of the minimal model, allowing, say, for  some generational 
non universality  in the soft supersymmetry breaking parameters
or for  an approximate R parity symmetry, would introduce 
several new parameters, with a non trivial structure on quarks and leptons
flavors which could accommodate   extra CP violating phases.
As is known, high energy supercolliders  are expected to provide for
precision determinations  of these supersymmetry parameters.
Regarding the  much studied sleptons pair production reaction 
\cite{fujii,peskin2,thomas}, one can define
a simple spin-independent  CP asymmetry observable
in terms of the difference of  integrated rates, $(\s_{JJ'}-\s_{J'J})$,
with  $\s_{JJ'} = \s (e^-+e^+\to \tilde e^-_J +\tilde e^+_{J'}) $, 
for the  case of sleptons pairs  of different flavors, $J\ne J'$.
Recent works,  based on  the mechanism of sleptons flavor oscillations, have
examined  for correlated slepton pairs production,
the flavor non-diagonal rates \cite{krasn,arkani1}
and the CP-odd flavor  asymmetries,  defined as, ${\cal A}_{JJ'}=
{\s_{JJ'} -\s_{J'J} \over \s_{JJ'} +\s_{J'J}  } $ \cite{arkani2,bowser}.  
Encouraging values of  order,   ${\cal A}_{JJ'} \approx 10^{-3} $  were predicted  at the  next linear colliders (NLC) energies \cite{arkani2,bowser}.
While the rates, $\s _{JJ'}$, depend on pairwise non-degeneracies  in the
sleptons mass spectra, the asymmetries, ${\cal A}_{JJ'}$, entail the 
much stricter conditions  that both non-degeneracies and mixing angles 
between all slepton flavors, as well as the CP odd phase, must not vanish.
 
Our main  observation in this work  is that the
R parity odd interactions  could  provide an  alternative mechanism
for explaining  flavor non-diagonal CP asymmetries through possible
complex CP odd phases incorporated in 
the   relevant dimensionless coupling constants.
  While  these  interactions can contribute 
to flavor changing changing processes already  at tree  level, 
their  contributions to CP asymmetries  involve
interference  terms between tree and loop amplitudes.
Two important questions then are, first, whether  the contributions from the 
RPV  (R parity violating) interactions, given the known bounds on the R parity odd coupling constants,
 could  lead to observable production rates; second,   whether the CP  asymmetries could reach observable levels. 
We shall present in this work  a study of 
the contributions to the CP asymmetries, in
the reactions, $e^-+e^+\to \tilde  e_{HJ} +\tilde e^{'\star }_{HJ'} ,
 \ [H=L, R, \ J\ne J']$,
at the high energy  leptonic colliders,  for center of mass energies
from the Z-pole up to $1000 $ GeV.   
The RPV  lepton number violating interactions are defined by the familiar superpotential, 
$W_{R-odd}=\sum_{ijk}[ \ud \l _{ijk} L_iL_j E^c_k+
\l ' _{ijk} Q_i L_j D^c_k ]. $
A comparison with the  oscillations mechanism should enhance 
the impact of  future experimental 
measurements of these  observables at  the future high energy colliders. 

The contents are organized into 3 sections. In Section \ref{secs3}, 
we develop the basic formalism for describing 
the scattering  amplitudes 
at tree and one-loop levels for the production of 
slepton pairs, $\tilde e^-_{L} \tilde e^+_L $ and  
$\tilde e^-_R \tilde e^+_R $.
In Section  \ref{secs4}, we present and discuss  our numerical results
for the integrated cross sections and  the CP asymmetries.
\section{Production of charged sleptons  pairs}
\label{secs3}
\subsection{General formalism}
\label{subsecs31}
\begin{figure} [h]
\begin{center}
\leavevmode
\psfig{figure=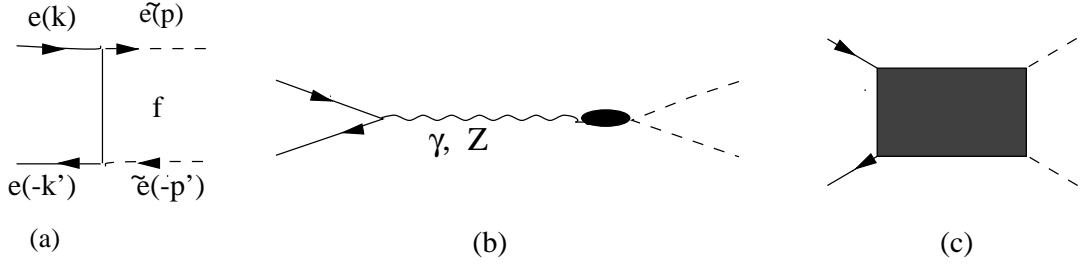}
\end{center}
\caption{Flavor non-diagonal process of $e^-e^+$ production of 
a sfermion-antisfermion pairs, $e^-(k)+e^+(k')\to \tilde e^-_J (p)
+\tilde  e^+_{J'} (p') $. The tree level diagram in $(a)$ represents
a neutrino, $f=\nu , \   t$-channel exchange amplitude. 
The  loop   level diagram in $(b)$ represents 
$\g -$  and $ Z- $ boson exchange amplitudes  with dressed vertices
 and that  in $(c)$ box amplitudes.  }
\label{figs1}
\end{figure}

The evaluation of spin-independent CP asymmetries  in  the production of a pair of sleptons, 
 $e^-(k)+e^+(k')\to \tilde e^-_{HJ}(p)+\tilde e^+_{H' J'} (p')$,  of different 
flavors, $J\ne J'$, with chiralities, $H=(L,R),\  H'=(L,R)   $, 
involves both tree and loop amplitudes.
Let us start  with the  case of  two left-chirality  sleptons, $H=H'=L$. 
At tree level, the R parity odd couplings, $\l _{ijk}$,   give
a non-vanishing contribution which is  
described by a  neutrino,  $\nu_i , \  t $-channel exchange Feynman 
diagram, as displayed  in  $(a)$ of Fig. \ref{figs1}. The associated flavor non-diagonal 
amplitude reads:
\begin{eqnarray}
M^{JJ'}_{tree} (\tilde e_L)= -{\l ^\star _{iJ1} \l _{iJ'1}\over t-m^2_{\nu_{i}}  }
\bar v(k') P_L(\kslash -\pslash )P_R u(k).
\label{eqs1}
\end{eqnarray}
Under our working assumption  that flavor changing effects are absent from  the 
supersymmetry breaking interactions, no other tree level  contributions  arise, 
since the gauge interactions can contribute, through the familiar
neutralinos $t$-channel  and gauge bosons $s$-channel exchanges, to flavor diagonal 
amplitudes, $J=J'$, only.

At one-loop level, there  occurs $\g -$ and  $Z-$boson 
exchange amplitudes with dressed $\g \tilde f\tilde {f}'$  and 
$Z \tilde f\tilde {f}'$  vertices involving three-point  vertex correction 
loop diagrams, as well as  box diagrams, of the type depicted
schematically in  $(b)$ and $(c)$  of  Fig.\ref{figs1}. 
We shall restrict consideration to the one-loop triangle diagrams 
contributions in the gauge bosons exchange  amplitude only.
Defining the dressed vertex functions for the  Z-boson
coupling to sleptons of chirality  $Z_\mu (P)\to 
\tilde f^J_H(p) +\tilde 
f^{J'\star }_H (p'), \ [H=L,R]$,  by the  effective Lagrangian,
\begin{eqnarray}
 L=-{g\over 2\cos \t_W} Z^\mu \G_\mu ^Z (p,p'), \quad 
\G_\mu ^Z(p,p')= (p-p')_\mu    [a_H(\tilde f_H) \d_{JJ'} 
+A^{JJ'}_H (\tilde f, s +i\e ) ] ,
\label{eqs2}
\end{eqnarray}
where, $ a(\tilde f_H)= a(f_H)=a_H(f)= 2T_3^H(f)-2Q(f) x_W, \ [x_W=\sin ^2\t_W]$ 
such that, $a(e_L)= -1+2x_W,\ a(e_R)= 2x_W$, 
we can express the one-loop Z-boson exchange amplitude as:
\begin{eqnarray}
M^{JJ'}_{loop}(\tilde e_H)&=&\bigg ({g\over 2\cos \t_W}\bigg )^2
\bar v(k')\g^\mu \bigg  (a(e_L)P_L+ 
a(e_R)P_R \bigg ) u(k) {1\over s-m_Z^2+im_Z\G_Z } \cr  &\times & (p-p')_\mu 
[a(\tilde e_H)\d_{JJ'}+ A_H^{JJ'} (\tilde e, s+i\e )],
\label{eq14}
\end{eqnarray}
where the shifted complex  argument,  $s+ i\e $, is  incorporated to remind us
 that  the vertex functions 
are complex functions in the complex plane of the 
Z-boson  virtual mass squared,  $s = (k+k')^2=(p+p')^2 $,  to be evaluated at 
the upper lip of the cut along the positive real axis.
In  the dressed  vertex function descibing the coupling,
$Z \tilde f \tilde f^\star $, eq.(\ref{eqs2}), 
we  have omitted the Lorentz covariant proportional to, $P_\mu =(p+p')_\mu 
=(k+k')_\mu $,  since this will give negligibly small lepton mass terms 
upon contraction in the total Z-boson exchange
amplitude, eq.(\ref{eq14}),  with the initial state leptons vertex covariant.
It is most convenient to describe the initial leptons polarizations 
in the helicity eigenvalue basis. 
In the limit of vanishing initial leptons masses, only the two helicity flip
configurations, $e^-_R e^+_L, \ e^-_L e^+_R,$ are non vanishing. While the gauge
bosons $s$-channel  exchange  contributes  to both of these configurations, the
R parity violating neutrino  $s$-channel exchange contributes only to the first. 
The summed tree and loop amplitude, $M^{JJ'}(\tilde e_L) = M_{tree}^{JJ'}(\tilde e_L)
+ M_{loop} ^{JJ'}(\tilde e_L) $,    in the relevant configuration, namely,
$e^-_R +e^+_L= e^-(h=-\ud ) +e^+(\bar h = \ud ) $, reads:
\begin{eqnarray}
M^{JJ'}(\tilde e_L)&=&M(e^-_R+e^+_L\to \tilde e^-_{LJ}+\tilde e^+_{LJ'})= 
-\ud \b s \sin \t \bigg  [
{\l^\star _{iJ1} \l _{iJ'1}\over t-m^2_{\nu_i}} \cr & +&2\bigg (
{g\over 2\cos \t_W }\bigg )^2 {a(e_R) 
A_L^{JJ'} (\tilde e , s+i\e ) \over s-m_Z^2+im_Z\G_Z  } \bigg ].
\label{eq15}
\end{eqnarray}
The Z-boson exchange contribution to the 
other helicity flip configuration, $e^-_L e^+_R$, is simply obtained by the
substitution, $ a(e_R)\to a(e_L)$. 
We also note that the $\g $ exchange contribution
has the same formal structure as that of the Z-boson exchange,
and can be easily  incorporated by adding to the above amplitudes 
the terms obtained by the replacements, 
${g\over 2 \cos \t _W } \to { g \sin \t_W \over 2},
\ a_{L,R}(f)\to 2Q(f), \ (s-m_Z^2  +im_Z \G_Z)^{-1}\to s^{-1},$ along 
with the substitution of Z-boson by photon vertex functions, 
$ A_{L,R}^{JJ'} (\tilde e, s+i\e ) \to A_{L,R}^{\g JJ'} (\tilde e, s+i\e )$.
The kinematical notations here refer to the center of mass system, where $\beta ={p\over k}=
{2p\over \sqrt s}, \ \theta $ is the scattering angle and
the differential cross section for unpolarized initial leptons reads: $d\s /d \cos \t = 
{p\over  128\pi s k } \sum_{pol} \vert M^{JJ'} \vert^2 .$
(For unpolarized beams, one must remove the polarization sums and multiply by a factor
of $4$. Our results agree with those quoted in \cite{peskin2}.)
Denoting the amplitude for the charge conjugate process, $e^- +e^+\to 
\tilde e^-_{HJ'} +\tilde e^+_{HJ}, \ [H=L,R]$,  by $\bar M^{JJ'} (\tilde e_H)$ 
and using the simple relationship, $\bar M^{JJ'} (\tilde e_H) 
= M^{J'J} (\tilde e_H) $, one can describe the decomposition into
tree and loop components for the pair of CP conjugate processes as, 
\begin{eqnarray}
M^{JJ'}(\tilde e_H)=a_0^{JJ'}+\sum_\a a_\a ^{JJ'}  F^{JJ'}_\a (s+i\e ), \ \ 
\bar M^{JJ'} (\tilde e_H) = a^{JJ' \star } _0+\sum_\a a^{JJ'\star } _\a  
F^{JJ'} _\a (s+i\e ).
\label{eqstt}
\end{eqnarray}
A spin-independent CP asymmetry 
can be defined in  the familiar way as the normalized difference of rates,
\begin{eqnarray}
{\cal A}_{JJ'}(\tilde e_H)= {\vert M^{JJ'}(\tilde e_H) \vert ^2- \vert \bar  M^{JJ'}
(\tilde e_H) \vert ^2\over 
\vert M^{JJ'}(\tilde e_H) \vert ^2+ \vert \bar  M^{JJ'} (\tilde e_H) \vert ^2 } 
\simeq {2\over \vert a_0\vert ^2}
\sum_\a  Im(a_0a_\a ^\star ) Im (F_\a (s+i\e )),  
\label{eq17}
\end{eqnarray}
where we have assumed  in the second step that the tree level flavor 
non-diagonal amplitude, $a_0$, 
dominates over the loop level amplitude, $a_\a  F_\a $,
and used the index  $\a $ to label the internal states running inside the loop.
\subsection{Loop amplitudes}
\label{subsecs32}
\begin{figure} [h]
\begin{center}
\leavevmode
\psfig{figure=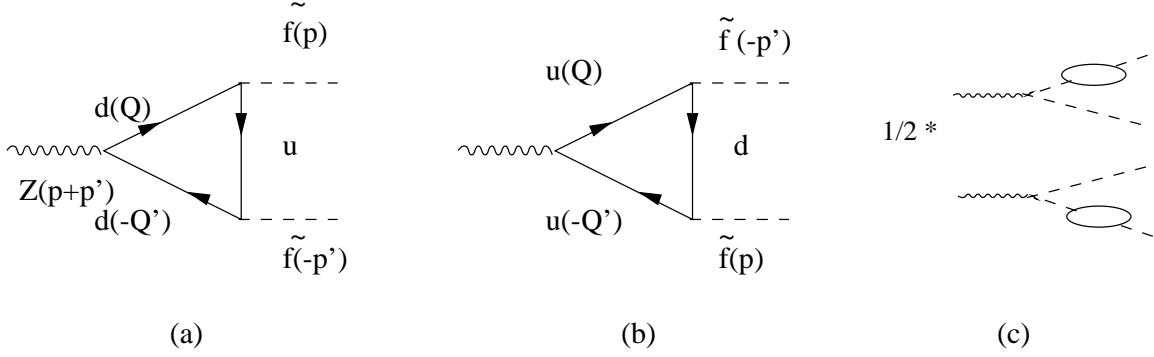}
\end{center}
\vskip 0.5cm
\caption{One-loop diagrams for the dressed $Z\tilde f \tilde f^\star $ vertex.
The flow of four-momenta for the intermediate  fermions  is  denoted  as, 
$ Z (P=k+k') \to f(Q)+\bar f(Q') \to \tilde f_J(p) + \tilde
f^\star _{J'} (p') $.  }
\label{figs2}
\vskip 0.5cm
\end{figure}

The  one-loop  triangle  diagrams,  describing the dressed vertex functions,
$Z\tilde f_L \tilde f_L^{\star } $, arise 
in two distinct charge configurations,   
shown in Fig. \ref{figs2}  by the diagrams $(a) $ and $(b)$, which involve
the d- and u-quark Z-boson  currents, respectively. 
The associated vertex functions read:
\begin{eqnarray}
&&\G_{\mu } ^Z (p,p')\vert _{a}=-i N_c{\l ' }_{Jjk}^\star 
\l ' _{J'jk} \cr  
&\times &\int_Q {  Tr[P_R(\qslash +m_{d_k}) \g_\mu 
(a(d_L)P_L+a(d_R)P_R)(-\Pslash +\qslash +m_{d_k})
P_L(\qslash -\pslash +m_{u_j  }  )]
\over (-Q^2+m_{d_k}^2)(-(Q-p-p')^2+m_{d_k}^2)(-(Q-p)^2+m_{u_j}^2)  }\ , \cr 
&&\G_{\mu } ^Z(p,p')\vert _{b}=-iN_c{\l ' }_{Jjk}^\star \l ' _{J'jk} 
\cr  & \times & 
\int_Q {  Tr[P_R(\qslash  +\pslash +m_{d_k})  P_L (\qslash +\pslash +\pslash '
+m_{u_j} ) \g_\mu (a(u_L)P_L+a(u_R)P_R)(\qslash +m_{u_j})  ]
\over (-Q^2+m_{u_j}^2)(-(Q+p+p')^2+m_{u_j}^2)(-(Q+p)^2+m_{d_k}^2)  }\ . \cr   &&
\label{eq12}
\end{eqnarray}
Applying the formalism  of Passarino-Veltman \cite{pasvel}, 
the vertex function from diagram $(a)$   can be expressed in the form:
\begin{eqnarray}
A_L^{JJ'}\vert_a  &=& 
{ {\lambda '}_{Jjk}^\star {\lambda '} _{J'jk}\over 2 (4\pi )^2 } 
N_c \bigg [2 a(d_L)m_d^2(C_0+C_{11}-C_{12})
+a(d_R)\bigg ( B_0^{(2)} +B_0^{(3)}+ 2 P\cdot p (C_{11}-C_{12}) \cr
&+&P^2C_0+2m^2_{\tilde e_J} (-C_{11}+C_{12}) 
-2m_d^2C_0+2 m^2_u(C_{11}-C_{12}) \bigg )\bigg ] \ .
\label{eq13}
\end{eqnarray}
The  conventions of ref.\cite{pasvel} are used for the  two-point and three-point 
integral functions, $ B_X \ [X=0,1] $ and $  \ C_X \  [X=0, 11, 12, 21,22,23,24]$. For notational convenience, 
  we have introduced the following abbreviations for the dependence on 
 argument variables:
$B_X^{(1)}= B_X(-p-p', m_d, m_d), \ B_X^{(2)}= B_X(-p , m_d, m_u),
\ B_X ^{(3)}= B_X(-p', m_u , m_d) $ and $C_X(-p,-p',m_d,m_u , m_d). $ 
 The amplitude from diagram $(b)$ can be 
 obtained from that of diagram $(a)$  by performing 
 the following   substitutions:  $ m_{d_k}\to m_{u_j}, \ p\to p', P_L\to P_R, 
 a(d_H) P_H \to a(u_H) P_H, \ [H=L,R]$. 
The self-energy contributions, which are represented 
by the diagrams $(c) $ in Fig. \ref{figs2},
with a single configuration only for the 
 d- and u-quarks  which propagate  inside the loop, 
are most conveniently  calculated through  a consideration of
the scalar  fields renormalization factors $Z_{JJ'}  $.
Starting from the schematic equations for the scalar field
$\phi $  bare Lagrangian density, 
 $L=\phi^\star (p^2-m^2+\Pi(p))\phi , $
where, $\Pi (p)= \Pi_1 p^2 -m^2 \Pi_0+\cdots  $, 
one transfers from bare   to 
renormalized quantities  by applying the substitutions,  
$\phi \to \phi/(1+\Pi_1)^\ud, m^2\to m^2(1+\Pi_1)/(1+\Pi_0)$, such that the renormalization equations for
the fields and mass parameters  read,  $\phi_J= Z_{JJ'} \phi^{ren}_{J'}, \ 
 m^2_{JJ'}=  Z^m_{JK} m^{ren 2}_{KJ'}  $,  with
$Z=(1+\Pi_1)^{-1 },\   Z^m = (1+\Pi_0)(1+\Pi_1)^{-1}$, using a matrix notation 
for the flavor dependence. The 
self-energy contribution in the vertex function becomes then,
\begin{equation} 
A_L^{JJ'}\vert_{SE}= [(Z_{JJ'} Z^\star _{JJ'})^\ud -1] \G_\mu ^Z =
2 N_c 
{ {\lambda '}_{Jjk}^\star {\lambda '} _{J'jk}\over (4\pi )^2 } 
a_L(\tilde e) B_1^{(2)}.
\label{eqself}
\end{equation}
Grouping together  the self-energy and the fermionic triangle   diagram contributions,
such that the total amplitudes read as, 
$A_L^{JJ'} (\tilde e)  =A_L^{JJ'} (\tilde e)_a  + A_L^{JJ'} (\tilde e)_b  $,
yields the final formulas:
\begin{eqnarray}
A_L^{JJ'}(\tilde e)_a&=&{N_c\over 2} 
{ {\lambda '}_{Jjk}^\star {\lambda '} _{J'jk}\over (4\pi )^2 } 
\bigg [2a(d_L)m_d^2 (C_0+C_{11}-C_{12} ) +
a(d_R) \bigg (B_0^{(2)}+B_0^{(3)}+2P\cdot p (C_{11}-C_{12}) \cr 
&+&P^2 C_0
+2m^2_J (-C_{11}+C_{12})-2m^2_dC_0+2m^2_u(C_{11}-C_{12}) \bigg )
+2a(\tilde e_L) B_1^{(2)} \bigg ] ,   \cr
A_L^{JJ'}(\tilde e)_b&=&-{N_c\over 2} 
{ {\lambda '}_{Jjk}^\star {\lambda '} _{J'jk}\over (4\pi )^2 } 
\bigg [2a(u_R)m_u^2 (C_0+C_{11}-C_{12} ) +
a(u_L) \bigg (B_0^{(2)}+B_0^{(3)}+2P\cdot p (C_{11}-C_{12}) \cr 
&+&P^2 C_0
+2m^2_{J} (-C_{11}+C_{12})-2m^2_uC_0+2m^2_d(C_{11}-C_{12})\bigg )
-2a(\tilde e_L) B_1^{(2)}
\bigg ]. 
\label{eq16}
\end{eqnarray}
For notational convenience, we have split the self-energy contribution 
into two equal parts that we absorbed  within the  above two amplitudes, 
distinguished by the suffices $a $ and $  b$. Note that the arguments 
in the  $B$- and $C$-integrals for
the  amplitude $ b$  are deduced from those  of the amplitude $a$   by replacing, 
$d_k \to u_j$. To obtain these results we have used 
the mathematica routine package  ``Tracer" \cite{jamin} and, 
for a cross-check, ``FeynCalc" \cite{calc}.
 A very useful check concerns the cancellation of the  ultraviolet  divergencies.
We indeed find that  the familiar \cite{pasvel} logarithmically  divergent term, $\Delta $, 
enters with the factors, $+a(\tilde e_L) -2a(d_R)$ (amplitude $a$) and 
$a(\tilde e_L) +2a(u_L) $  (amplitude $b$), whose total sum  vanishes 
identically.

The interactions associated with 
the  coupling constants, $\l_{ijk} $, can  also contribute at one-loop order.
Exploiting  the formal similarity between  the $ \l $ and $ \l '$  
interaction terms  in the Lagrangian density,
namely,   $L=-\l' _{ijk} \tilde e_{iL} \bar d_{kR} u_{jL} -  
\l_{ijk} \tilde e_{iL} \bar e_{kR} \nu_{jL} + \cdots $,  dispenses us from 
performing  a new calculation. The results can be derived from those in 
eq.(\ref{eq16}) by substituting   for  the internal lines,
$ d_k \to e_k, \  u_j \to \nu_j $, and for the parameters, 
$a_H(u)\to a_H(\nu ), \ a_H(d)\to a_H(e), \ \ 
 \l ^{'\star } _{Jjk} \l '_{J'jk} \to    \l ^{\star } _{Jjk} \l _{J'jk}  $. 

 Let us now turn to the production of right-chirality sleptons where 
analogous results  can be derived.
The tree level amplitude is related to that in eq.(\ref{eqs1}) 
by a simple  chirality change,
 \begin{eqnarray}
M^{JJ'}_{tree} (\tilde e_R)= -{\l ^\star _{i1J'} \l _{i1J}\over t-m^2_{\nu_{i}}  }
\bar v(k') P_R(\kslash -\pslash )P_L u(k).
\label{eqs17}
\end{eqnarray}
There occurs only one non-vanishing   helicity  flip 
configuration for the initial leptons, namely,  $e^-_L e^+_R$, 
in which the neutrinos  $ t$-channel and  the gauge bosons 
$s$-channel contributions 
interfere. The  amplitude is given by a  formula similar to eq.(\ref{eq15}), except
for  the substitution in the second term,
$ a_R(e) A_L^{JJ'} (\tilde e, s+i\e ) \to 
 a_L(e) A_R^{JJ'} (\tilde e, s+i\e )$. Concerning  the  one-loop contribution 
to  the vertex function  $A_R^{JJ'}(\tilde f)$, 
we find that the RPV interactions with the  coupling constants $\l_{ijk} $ 
can only  contribute, while those with $\l '_{ijk}$ vanish identically.
Diagram $(a)$ in Fig. \ref{figs2} refers to an $e_j $ current 
and diagram $(b)$ 
to a $\nu_i^c$ current.
The results can be derived by  inspection   from  eq.(\ref{eq16}) by substituting,  $ \l '_{J'jk} \l ^{'\star } _{Jjk} \to 
\l_{ijJ} \l^\star _{ijJ'}$,  
 $ d_{jR} \to e_{jL}, u_{jL} \to \nu_{iR}^c, \ \tilde e_L \to 
\tilde e_R $ and, 
accordingly,  $a(d_H)\to a(e_H), \ 
a(u_H)\to a(\nu^c_H), \ [H=L,R], \ a(\tilde e_L)  \to a(\tilde e_R)  $. 
For definiteness, we quote the explicit formulas:
\begin{eqnarray}
A_R^{JJ'}(\tilde e)_a&=&{N_c\over 2} 
{ {\lambda }_{ijJ} {\lambda }_{ijJ'}^\star \over (4\pi )^2 }
\bigg [2a(e_R)m_e^2 (C_0+C_{11}-C_{12} ) +
a(e_L) \bigg (B_0^{(2)}+B_0^{(3)}+2P\cdot p (C_{11}-C_{12}) \cr 
&+&P^2 C_0
+2m^2_J (-C_{11}+C_{12})-2m^2_eC_0+2m^2_{\nu }(C_{11}-C_{12}) \bigg )
+2a(\tilde e_R) B_1^{(2)} \bigg ],  \cr
A_R^{JJ'}(\tilde e)_b&=&-{N_c\over 2} 
{ {\lambda }_{ijJ} {\lambda }_{ijJ'}^\star \over (4\pi )^2 } 
\bigg [2a(\nu_L^c)m_{\nu }^2 (C_0+C_{11}-C_{12} ) +
a(\nu_R^c) \bigg (B_0^{(2)}+B_0^{(3)}+2P\cdot p (C_{11}-C_{12}) \cr 
&+&P^2 C_0
+2m^2_{J} (-C_{11}+C_{12})-2m^2_{\nu }C_0+2m^2_e(C_{11}-C_{12})\bigg )
-2a(\tilde e_R) B_1^{(2)}
\bigg ]. 
\label{eq18}
\end{eqnarray}
The discussion of the mixed chiralities cases, $\tilde e_{LJ}^-\tilde e_{RJ'}^+, \ 
\tilde e_{RJ}^-\tilde e_{LJ'}^+, \ [J\ne J']$ turns out to be quite brief.  
The tree level RPV contributions, which  can only come from the $\l_{ijk}$ interactions,
vanish identically for massless neutrinos. As for the one-loop
contributions to the vertex, $Z\tilde f_L\tilde f_R^\star $, this also vanishes up to mass 
terms in the internal fermions. Since flavor non-diagonal rates arise then from 
loop contributions only and CP asymmetries from  interference of distinct  loop
contributions, one concludes that both observables should be very small. 

Finally, let us add here a general comment  concerning 
the photon vertex functions, $ A_{L,R}^{\g JJ'} $, which  are  given by 
formulas  similar to those in eqs.(\ref{eq16}) or (\ref{eq18}) with the 
appropriate replacements, $a_{L,R}(f) \to  2 Q(f)$. Therefore, to incorporate
the $\g $-exchange contributions in the total
amplitudes (eq.(\ref{eq15}) and related equations)
one needs to substitute, 
$$ a_{R,L}(e) A^{JJ'}_{L,R} \to 
a_{R,L}(e) \sum_f a(f) C_f+ 2Q(e)  \sin^2 \t_W \cos^2 \t_W  
[(s-m_Z^2+im_Z \G_Z) / s] \sum_f 2Q(f) C_f, $$
where we have used the schematic representation, $A^{JJ'}_{L,R}=\sum_f a(f) C_f$.
\begin{figure} [h]
\begin{center}
\leavevmode
\psfig{figure=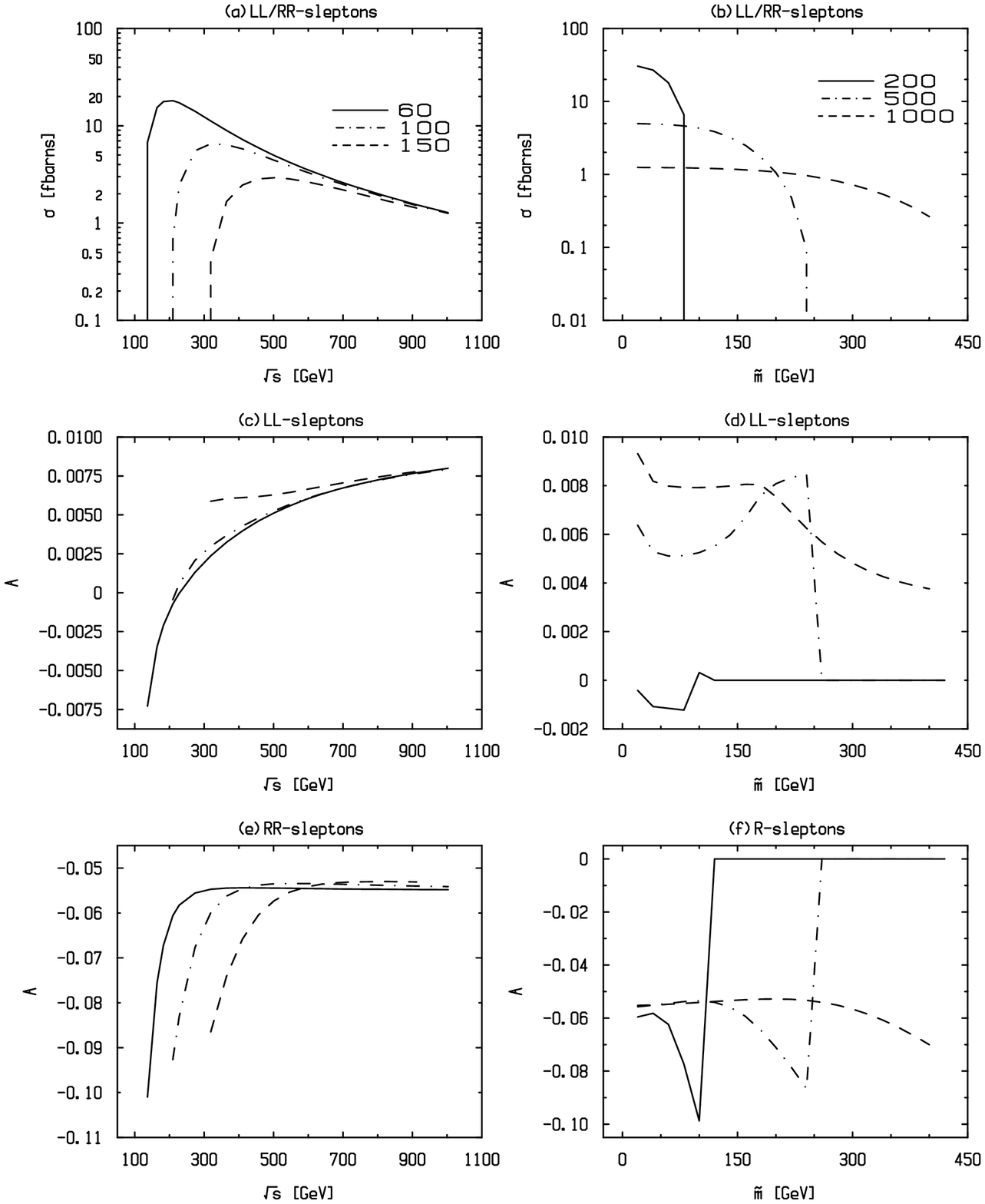}
\end{center}
\end{figure}
\begin{figure}
\caption{Integrated flavor non-diagonal cross sections and CP asymmetries  
in the production of  slepton-antislepton pairs  of left-chirality (L) 
(interactions $\l '_{ijk}$ only) and of  right-chirality  (R) 
(interactions $\l _{ijk}$). The  three windows on
 the left-hand side ($(a), \ (c), \ (e) $)  show the variation with center of mass
energy, $s^{1/2}$,  for three choices of the  scalar superpartners mass parameter,  $\tilde m $: 
$ 60 GeV$ (continuous lines), $ 100 GeV$ (dashed-dotted  lines),  
$  150 GeV$ (dashed lines).  The  three windows on the right-hand side  ($(b),\ (d),\ (f) $) 
show the variation with  scalar superpartner mass, $\tilde m$, 
for three  choices  of the center of mass energy
$  s^{1/2} = 200 GeV$ (continuous lines), $ 500 GeV$ (dashed-dotted  lines),
$  1000  GeV$ (dashed lines).  The tree level amplitude includes the  t-channel 
exchange contribution. The one-loop  amplitudes (with both  photon and Z-boson exchanges)  correspond to Case {\bf IV} which 
includes the contributions from all three internal fermions  generations.} 
\label{figg6}
\end{figure}
\section{Results and discussion}
\label{secs4}

Let us first comment briefly on the experimental  observability 
of flavor non-diagonal sleptons pair production.
One  convenient non degraded signal here  is that which corresponds
to lepton pair final states, $e_J ^-e^+_{J'} $,
which are produced through the  two-body decay channels for sleptons,
$\tilde e_{[J, J']} ^\pm 
\to e^\pm _{[J , J']} +\tilde \chi_1^0$.  Of course,  in the broken 
R parity case, the  produced lightest neutralinos  are unstable  
and could conceivably be  reconstructed 
through  their dominant  decay channels which involve
two leptons, or two jets, together with missing energy. 
We shall not elaborate further on this issue, except to note 
that the efficiency factors at NLC energies  for the flavor diagonal rates, assuming 
a stable $\tilde \chi_1^0$, and including rough detection cuts,
such that the physical rates  for the fermion pairs channels is, $\s _{JJ'} \e $,
are typically set at $\e \approx 30 $ \% \cite{arkani1}.

Proceeding to the predictions, we observe that
the  main source of uncertainties concerns the RPV coupling constants.
The sfermion mass eigenvalues are not known, but these
parameters appear explicitly  through the kinematics. 
We shall neglect mass splittings  and mixings  between L- and R-sleptons.
A  unique sleptons mass parameter, $\tilde m$,  will be used  and
varied in  the interval, 
$60 < \tilde m <  400$ GeV.   Regarding the RPV coupling constants,
it is useful  here to catalog the family configurations and intermediate states 
entering the calculations. 
Examining the structure of  the  flavor non-diagonal 
 tree amplitudes, we note  that these involve a onefold 
summation  over leptons  families weighted by the factors,
$ t^i_{JJ'}=  \l _{iJ1}^\star \l_{iJ'1}  $, for L-sleptons and  
$ t^i_{JJ'}=  \l _{i1J}^\star \l_{i1J'}  $, for R-sleptons. 
The loop amplitudes involve a twofold summation  over leptons families of form, 
$\sum_{jk} l^{jk}_{JJ'}  F^{jk} (m_j, m_k, s+i\e ) , \ $ where  $l^{jk}_{JJ'}$ 
depend quadratically on the RPV coupling constants while
the  loop integrals, $F^{jk} $, have a  non-trivial
dependence on the fermions    masses, as exhibited 
on the formulas  derived in Section \ref{secs3} 
[see, e.g., eq. (\ref{eq16})].  The relevant coupling constants, the 
species and family configurations  for the internal fermions are for L-sleptons,
$ l^{jk}_{JJ'}=  \l ^{'\star }_{Jjk}
\l ' _{J'jk} , 
[d_k, \ u_j]; \ \    
 l^{jk}_{JJ'}=  \l ^{\star }_{Jjk}
\l _{J'jk} , 
[e_k, \ \nu _j]$; 
and for R-sleptons,
$ l^{ij}_{JJ'}=  \l _{ijJ}
\l ^{\star } _{ijJ'} ,\  [e_j,\  
\nu ^c_{i}  ] $.
The dependence of rates on the RPV coupling constants 
has the schematic structure,  $ \s _{JJ'}\simeq \sum_i\vert t^i_{JJ'}\vert^2$, 
and that of CP asymmetries, 
$ {\cal A}_{JJ'} \simeq  \sum_{ij k} Im(l^{jk}_{JJ'} 
t^{i\star }_{JJ'})/\sum_l\vert t^l_{JJ'} \vert ^2$ for L-sleptons and
$ {\cal A}_{JJ'} \simeq  \sum_{ij k} Im(l^{ij}_{JJ'} 
t^{k\star }_{JJ'})/\sum_l\vert t^l_{JJ'} \vert ^2$ for R-sleptons. 
Therefore,  rates (asymmetries) are controlled by two (four) RPV
coupling constants in different family configurations. 
Note the  expected invariance of asymmetries under phase redefinitions of the fields.

While the dependence on the mass of the exchanged neutrino family index in 
$t^i_{JJ'}$  can be clearly  ignored, that on the pair  of indices $(i,j)$ in $l^{ij}_{JJ'}$, which involves  the ratios of the masses
of the appropriate internal  fermions, $m_{i,j} $, to the external scale associated with the
center of mass energy, $\sqrt {s}$, can be ignored as long as, 
$\sqrt {s} >> m_{i,j}$. Therefore,  at the energies of interest, the only relevant
fermion mass parameter is  that of the  top-quark. 
Instead of listing the various distinct family configurations 
for the quadratic (tree)  or quartic (loop) 
products of the RPV coupling constants, we shall consider a 
 set of specific assumptions concerning the family dependence.
First, for  the cases involving $[e_j, \nu_i^c]$ or $[e_k, \nu_j]$ internal states, neglecting neutrino masses, 
we  need only account for the masses of charged leptons. 
For the case  with $[d_k, \ u_j]$ internal states, 
we  restrict consideration to  the diagonal family configuration,
namely, $k= j $. Second, we  include a CP odd phase, $\psi $,
between $t^i_{JJ'}$ and  all of the $l^{jk}_{JJ'} $ or $l^{ij}_{JJ'}$, as the case may be. Finally, we consider the  following four 
 discrete  choices for the variation intervals on which 
run the internal fermion indices indices,
$j = k$ or $i, \ j$. 
Case {\bf I}:  $\{ 1 \}$;  Case {\bf II} $\{ 2 \}$;
 Case  {\bf III} $\{ 3 \}$;  Case  {\bf IV} $\{ 1, 2,3\}$. 
In all these four  cases, we set the relevant coupling constants 
at the reference values, $l^{ij}_{JJ'}  = l^{jk}_{JJ'}
=10^{-2},\  t^i_{JJ'}=10^{-2}$ and use a maximal CP odd phase,
 $ arg( l_{JJ'} ^{\star [ij, jk] } t^l_{JJ'} )  \equiv  \psi =\pi/2$. 
Because of the proportionality of asymmetries to  the  imaginary part of the phase factor, the  requisite dependence may be simply
reinstated by inserting a factor, $\sin \psi $. 
\begin{table}
\begin{center}
\caption{CP asymmetries, ${\cal A}_{JJ'}$, 
in sleptons pair production at two values of the 
center of mass energy, $s^{1/2}= 200, \ 500 $ GeV and for 
values of the sleptons mass parameter, $\tilde m = 60, \ 100, \ 200 $ GeV,
appearing in the column fields.  For each case, the first line  ($Z$) 
is associated with the  gauge Z-boson exchange contribution and the second
line ($\g +Z$) with both photon and Z-boson exchanges added in together.
The contributions  to left-chirality ($\tilde e_L\tilde e_L$) 
and right-chirality ($\tilde e_R\tilde e_R$) sleptons,  
induced by the $\l '_{ijk}$ and $\l _{ijk}$ interactions,  are distinguished 
by the labels, $\l ' \l ^{'\star } , \ \l  \l ^\star  $, respectively. 
Cases {\bf I , \ III } correspond to internal fermions belonging to 
the first  and  third families, respectively. 
The notation $ n d-x $ stands for $ n \  10^{-x}$. }
\vskip 0.3cm 
\begin{tabular}{cccccc}
\hline 
    & &   $s^{1/2} = 200 $ GeV &    $ s^{1/2} =500 $ GeV &    &   \\
\hline
 & &    $\tilde m =60 $ &      $\tilde m =60$ & 
  $\tilde m =100 $ &  $\tilde m =200$    \\
\hline
${\bf \tilde e_L \ \tilde e_L }$ \\
\\
$\lambda ' \lambda '^\star $  \\
    {\bf  I} &$ Z$ & $ -2.1 d-5     $& $-3.3 d-6    $& $-2.6 d-6     $& $-2.4d-6    $\\
 & $\g +Z$    & $ -7.7 d-5     $& $-1.39 d-5    $& $-1.09 d-5     $& $-1.03d-5    $\\
   \\
 {\bf  III}& $Z$ & $ +2.6 d-4      $& $-1.6 d-3    $& $ -1.8 d-3    $& $-2.3d-3    $\\
&   $\g +Z$ & $ -1.01  d-3      $& $+5.1 d-3    $& $ +5.3 d-3    $& $+8.1d-3    $\\
   \\
   \\
$\lambda  \lambda ^\star $  \\
    {\bf  I} &$Z$ & $ -2.1 d-5      $& $ -3.3 d-6    $& $-2.6d-6     $& $-2.4d-6    $\\
 &   $\g +Z$ & $ -7.69 d-5      $& $ -1.39 d-5    $& $-1.09d-5     $& $-1.03d-5    $\\
   \\
    {\bf  III} &$Z$ & $-2.4 d-5      $& $-5.5 d-6    $& $-3.4 d-6     $& $ -2.7 d-6   $\\
 &   $\g +Z $ & $-6.39 d-5      $& $+2.59 d-6    $& $-5.06 d-6     $& $ -8.32d-6   $\\
   \\
   \\
\hline
${\bf \tilde e_R\   \tilde e_R}$ \\
\\
$\lambda  \lambda ^\star $  \\
    {\bf  I} &$Z$ & $ -7.2 d-3      $& $-5.5 d-3    $& $-5.4 d-3     $& $ -7.2 d-3    $\\
&  $\g +Z$ & $ -2.1 d-2      $& $-1.83 d-2    $& $-1.80 d-2     $& $ -2.40 d-2    $\\
\\
\hline 
\end{tabular}
\label{table1}
\end{center}
\end{table}
To illustrate the dependence of asymmetries on the internal fermions families
and on the $\l '$ or $\l $  interaction  types, 
we display in Table \ref{table1}   a set of representative 
results obtained for selected subsets of Cases {\bf I, \ II , \ III, \ IV}. 
The reason is that the results for Cases {\bf I , \ II} (light families)
are identical in all cases, while those for Case {\bf III} (heavy 
families) differ only  for cases  involving up-quarks. 
As one sees on Table \ref{table1}, the  interference between photon and Z-boson  exchange 
contributions  has a significant  effect on the results.  
The  strongly reduced values for the L-sleptons  asymmetries found 
in Cases {\bf I} for the $\l ' \l'^\star $ interactions and in  
all Cases  for the $\l  \l^\star $ interactions, arise from 
the existence of a strong cancellation between  the amplitudes termed $(a)$
and $(b)$ for  nearly  massless  internal  quarks or leptons. 
Case {\bf III} with the $\l ' \l'^\star $ interactions 
is relatively enhanced thanks to the top-quark contribution (configuration 
$\bar t \ b$).  That the above  cancellation is  not 
generic  to the RPV contributions 
is verified on the results  for R-sleptons production, 
 where all three  families of leptons give nearly 
equal, unsuppressed  contributions to loop amplitudes.

In the currently favored  situation  where the RPV
 coupling constants  are assumed to exhibit a strong 
hierarchical structure, the peculiar rational 
dependence of CP asymmetries on ratios of
quartic products of the coupling constants, might  lead to strong enhancement
factors.  We recall the schematical structure of this dependence,
${\cal A}_{JJ'}\propto  [ \sum_{ijk}  Im (\l ^{'\star } _{Jjk} \l '_{J'jk} \l '_{iJ1} \l ^{'\star } _{iJ'1} ) 
/ \sum _l \vert  \l _{lJ1} \l ^\star _{lJ'1}\vert ^2   ] $, 
and note that the coupling constants involving  third family  indices are 
amongst those that are the least  strongly constrained. Therefore,  assuming 
that the coupling constants take the values given by the 
current bounds from low energy constraints \cite{reviews}, 
one  would  obtain,  
$${\cal A}_{J=3,J'=2} 
\simeq [ Im (\l ^{' \star }_{333} \l '_{323} \l '_{331} \l ^{'\star }_{321})
/ \vert \l _{131} \l _{121} ^{'\star }  \vert ^2 ]\approx 90 \sin \psi  .$$

The  dependence  of rates and asymmetries on center of mass energy and
sleptons masses are displayed  for Case {\bf IV} in 
Figure \ref{figg6}. Regarding the variation with energy (figure (a)), 
after a rapid rise at  threshold (with the expected $\b ^3  \ p$-wave like
behavior)  the rates settle, roughly as $\tilde m^2/s$, 
to constant values with growing energy,   and
vary  inside the range, $ ({\l ^\star \l \over 0.01} )^2 \ 20 \ - \
2$ fbarns,  as one sweeps through the interval,   $\tilde m \in [60, 400]$ GeV.
The variation with $\tilde m$ (figure (b))  is rather smooth.
For the envisaged integrated luminosities,
 $ {\cal L} \simeq 50 \ - \ 100  
fbarns ^{-1}  /yr $,  these results indicate that reasonably sized samples of 
order $100 $ events could be collected at NLC.  
Noting that the dependence of rates on energy
 rapidly saturates for $\sqrt s > \tilde m$, we conclude that the
relevant  bounds that could be inferred on quadratic  products
of different  the RPV coupling constants, should, for increasing sleptons 
masses, become competitive   with those deduced from 
low energy constraints, which scale typically as,
$[\l   \l  , \ \l '  \l ' ] <  0.1 (100 GeV /\tilde m)^2$. 
The results in Fig.\ref{figg6} (c,d,e,f) for the  CP  asymmetries,
${\cal A}_{JJ'}$,   indicate the existence of a wide, nearly 
one order of magnitude, gap between L-sleptons with 
$\l ^{' \star } \l '$ interactions 
 and R-sleptons with  $\l ^{ \star } \l $ interactions,  with values that lie 
at a few times $10^{-3}$  and $10^{-2}$, respectively.

In our  prescription of using  equal numerical  values for the RPV coupling constants ($t^i _{JJ'} $ and $l^{ij}_{JJ'}$)  which
control tree and loop contributions,
the asymmetries are independent of the specific reference values chosen.
In the event that the rates would be dominated by some alternative mechanism,
say, lepton flavor  ocillations, whereas RPV effects would 
remain significant in
asymmetries, these would then  scale as, $Im(t^{i\star}_{JJ'} l^{jk}_{JJ'})$.  
It is instructive in view of such a possibility  to compare with 
predictions  found in  the flavors  oscillation approach.
Scanning over wide intervals of values for  the  relevant  parameters,
$[\cos 2\t_R , x=\D \tilde m/\G ] $, 
associated with the common values for all three  
mixing angles  and ratios  of  families mass  differences  to the total
sleptons decay widths, respectively,  the authors of \cite{arkani1} found 
flavor non-diagonal rates  which ranged between 
$ 250  $ and $0.1 $ fbarns for $ \sqrt s =190 $ GeV and 
$ 100 $ and $ 0.01 $ fbarns for $\sqrt s =500 $ GeV.
Our predictions,  $ \s _{JJ'} \simeq ({\l \over 0.1 })^4 \  2 \ - 20 $ fbarns, 
which hold  approximately for energies, $\sqrt  s > \tilde m$, 
lie roughly in between these extreme values. 
On the other hand, the authors of \cite{arkani2}
found CP  asymmetry rates, $S_{JJ'} =
\s_{JJ'}-\s _{J'J} \approx  3-16 $ fbarns.  
For comparison,  our predicted asymmetry rates for the same quantity, 
namely,  $S_{JJ'}= 2\s_{JJ'} {\cal A}_{JJ'}
\approx 10^{0} - 10^{-1}$ fbarns, lie around one order of magnitude 
below these values.
It should be said, however,  that  the flavor oscillation contributions
could have a stronger model dependence  than  the 
variation range exhibited by the above predictions, and that 
these  predictions were 
obtained subject to assumptions that tend to maximize CP violation effects.
The existing constraints, \cite{fcnc} which are mostly derived from low energy
phenomenology, constrain only a small subset of the parameters
describing the scalar superpartners  mass spectra and generational mixings.

To summarize, we have shown that moderately small contributions to 
flavor non-diagonal  rates  and CP violating  spin-independent asymmetries 
in sleptons pair production  could arise from the RPV interactions. 
These contributions  seem to be of smaller 
size than those  currently associated with flavor oscillations, 
although  the model dependence of predictions  
in the flavor oscillation approach is  far from being under control.
An experimental observation of the  non-diagonal  slepton production rates 
would give information on  quadratic products of
different coupling constants, $\l \l^\star $. Owing to the smooth dependence 
of rates on the slepton masses, already for  masses, $\tilde m > 100 $ GeV,  
it should be possible  here to deduce  stronger bounds than  the current ones inferred from 
low energy constraints. The  observation of CP  violating 
asymmetries  requires  the presence of non vanishing CP odd phases in 
quartic products of the coupling constants,
$ Im (\l ^{\star } _{Jjk} \l _{J'jk} \l _{iJ1} \l ^{\star } _{iJ'1} ) $, 
(and similarly with $\l \to \l '$)
which remain largely unconstrained so far.  The peculiar rational dependence, 
$Im (\l \l ^\star  \l \l  ^\star )/\l ^4$, leaves room for possible
strong enhancement factors.

\end{document}